\documentclass[12pt]{article}
\usepackage{amsmath,amssymb,graphicx,multicol}
\def\be{\begin{equation}}
\def\ee{\end{equation}}
\def\ba{\begin{eqnarray}}
\def\ea{\end{eqnarray}}

\def\bdm{\begin{displaymath}}
\def\edm{\end{displaymath}}
\def\la{~\mbox{\raisebox{-.6ex}{$\stackrel{<}{\sim}$}}~}

\def\bq{\begin{quote}}
\def\eq{\end{quote}}

 at 10truept

\newcommand{\beq}{\begin{equation}}
\newcommand{\eeq}{\end{equation}}
\newcommand{\beqa}{\begin{eqnarray}}
\newcommand{\eeqa}{\end{eqnarray}}

\def\la{~\mbox{\raisebox{-.6ex}{$\stackrel{<}{\sim}$}}~}

\def\ltap{\ \raise.3ex\hbox{$<$\kern-.75em\lower1ex\hbox{$\sim$}}\ }
\def\gtap{\ \raise.3ex\hbox{$>$\kern-.75em\lower1ex\hbox{$\sim$}}\ }
\def\gl{\ \raise.5ex\hbox{$>$}\kern-.8em\lower.5ex\hbox{$<$}\ }
\def\roughly#1{\raise.3ex\hbox{$#1$\kern-.75em\lower1ex\hbox{$\sim$}}}

\begin{document}

\thispagestyle{empty}
\begin{flushright}
May 2007
\end{flushright}
\vspace*{1.2cm}
\begin{center}
{\Large \bf How Black Holes Form in High Energy Collisions}\\

\vspace*{2cm} {\large Nemanja Kaloper\footnote{\tt
kaloper@physics.ucdavis.edu} and John Terning\footnote{\tt
terning@physics.ucdavis.edu} }\\
\vspace{.5cm} {\em Department of Physics, University of
California, Davis,
CA 95616}\\
\vspace{.15cm} \vspace{1.5cm} ABSTRACT
\end{center}

We elucidate how black holes form in trans-Planckian collisions.
In the rest frame of one of the incident particles, the
gravitational field of the other, which is rapidly moving, looks
like a gravitational shock wave. The shock wave focuses the target
particle down to a much smaller impact parameter. In turn, the
gravitational field of the target particle captures the projectile
when the resultant impact parameter is smaller than its own
Schwarzschild radius, forming a black hole. One can deduce this by
referring to the original argument of escape velocities exceeding
the speed of light, which Michell and Laplace used to discover the
existence of black holes.

\vskip4.5cm


\vfill \setcounter{page}{0} \setcounter{footnote}{0}
\newpage

Quantum gravity has long been a holy grail of fundamental theory.
The quest for it has been slow and arduous, in no small part due
to an acute lack of experimental milestones that could point the
way. Until recently this seemed inevitable, due to the unbearable
feebleness of gravity. However the advent of the braneworld
paradigm has rekindled some hopes that nature might be kinder to
us. Perhaps gravity is stronger at sub-mm distances, giving the
prospects that in some future collider experiments effects of
quantum gravity might be directly encountered. If the fundamental
scale of gravity is really low, near a ${\rm TeV}$, such effects
might be within the reach of LHC \cite{add}. Among the possible
direct signatures, certainly the most dramatic phenomena may be
the formation of black holes in high energy collisions, and their
subsequent decay by Hawking radiation. This interesting
possibility has attracted a lot of attention from theorists and
experimentalists alike, spurring a great deal of exploration of
the likelihood and signatures of such processes, as exemplified in
\cite{bhlhc}.

The exact description of black hole formation in a collision of
real particles does not yet exist. As a result, the analyses of
black hole formation processes \cite{bhlhc} estimate the black
hole production cross section by the horizon area of a black hole
whose horizon radius $r_h$ is set by the center-of-mass collision
energy $\sqrt{ s}$. When the impact parameter ${ b}$ is smaller
than $r_h$ it is assumed that the efficiency of formation of a
black hole is close to $100\%$. The strongest support for this
comes from examining the gravitational shock wave fields of zero
rest mass particles \cite{penrose,deathpayne,eardley}, and seeing
when a trapped surface forms as the shock waves cross. These
analyses, in particular the work of Eardley and Giddings
\cite{eardley} which covers off-center collisions with ${ b} > 0$,
find that trapped surfaces form when ${ b} \la r_h$, and have the
area scaling as the horizon area $\sim r_h^2\sim G_N^2 { s}$,
where $r_h$ is the horizon radius and $G_N$ is Newton's constant.
However the shock wave solutions inevitably break down when the
fields of different particles cross. Since the black hole cross
sections are extracted from the shocks, {\it a priori} they remain
sensitive to the nonlinear corrections, which still obscures the
interpretation of the results.

On the other hand, the cross section estimates by the horizon area
for off-center collisions were disputed in \cite{voloshin}. In
these papers, Voloshin has eloquently argued that the black hole
formation cross section should be exponentially suppressed for
off-center collisions, by the exponent of the Euclidean action of
the black hole. In a nutshell, the main tenet of this argument was
that at very high energies the direct local interactions of
particles must be strongly suppressed because of Lorentz boosts.
At high speeds, the particles occupy small regions of space
$\lambda \sim 1/p \ll { b}$, in the units $c = \hbar = 1$. Thus,
the argument goes, they don't get to interact strongly unless the
collisions are head-on, and so black hole formation cannot be a
local process. Instead, contends \cite{voloshin}, the description
of black hole formation must be based on some kind of instanton
solution, which would give an exponential suppression whenever ${
b} > 0$.

Given these doubts one might be led to consider an extreme case: a
frame of reference where one particle is almost at rest and the
other is highly boosted. The highly boosted particle is nothing
like a black hole and the slowly moving particle at a large impact
parameter is a tiny perturbation. So, how do black holes really
form in collisions? We will describe a very simple and intuitively
clear physical picture whereby a black hole forms in a {\it
classical capture} process. This is why the estimate of black hole
formation cross section by the classical horizon area is the
correct thing to do, as argued by
\cite{penrose,deathpayne,eardley}, and why the nonlinear
corrections should {\it not} change it qualitatively. To see this,
it is enough to use the original, simple escape velocities
argument, that led John Michell \cite{michell} in 1784 and Pierre
Simon, Marquis de Laplace \cite{laplace} in 1796 to deduce the
existence of black holes, some 130 years before the inception of
General Relativity. The details of trapped surface formation are
not essential here, although they may be necessary for comparison
with precision black hole experiments.

\begin{figure}[t]
\centering a)
\includegraphics[width=0.4\textwidth]{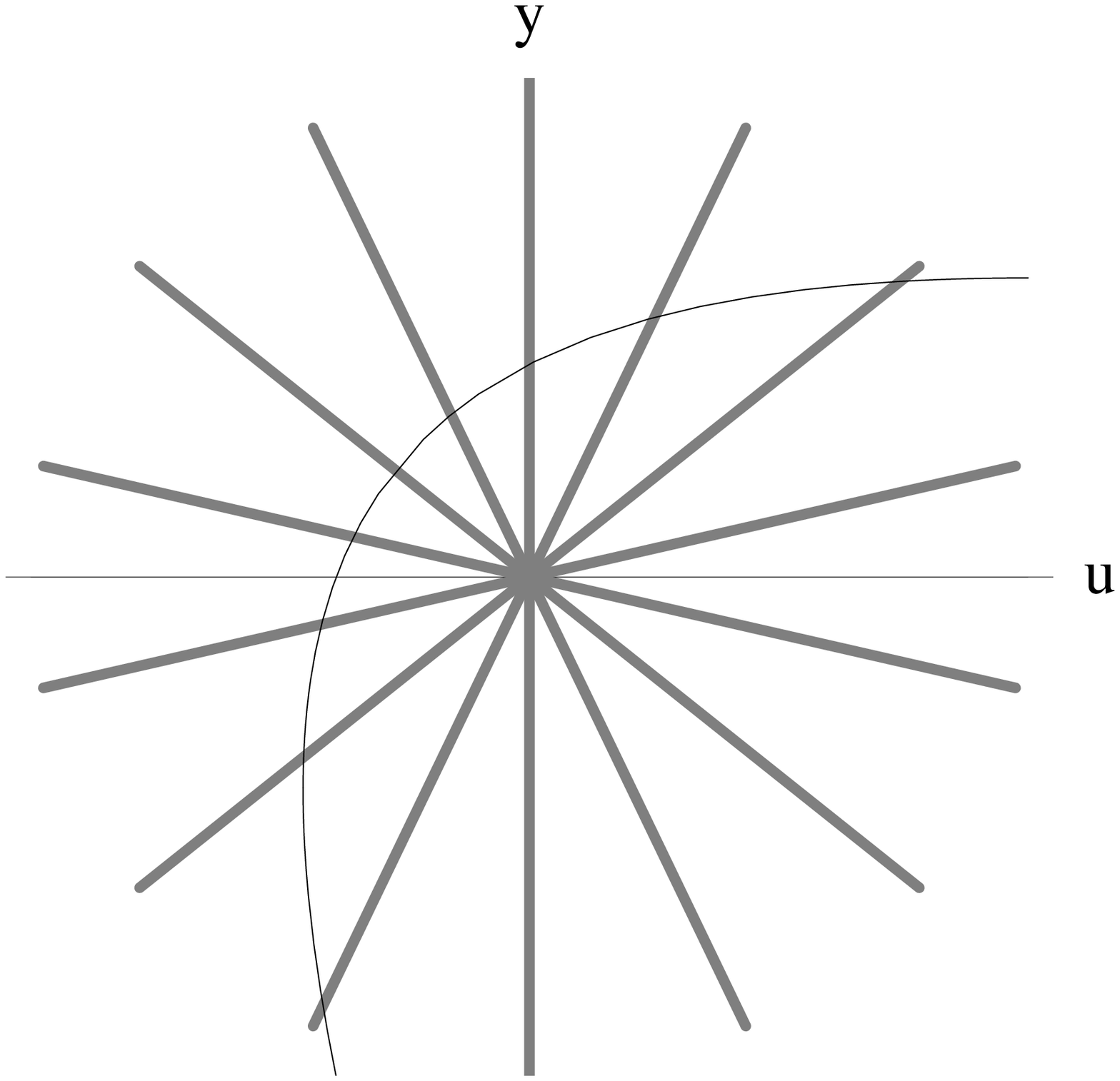}
\centering b)
\includegraphics[width=0.4\textwidth]{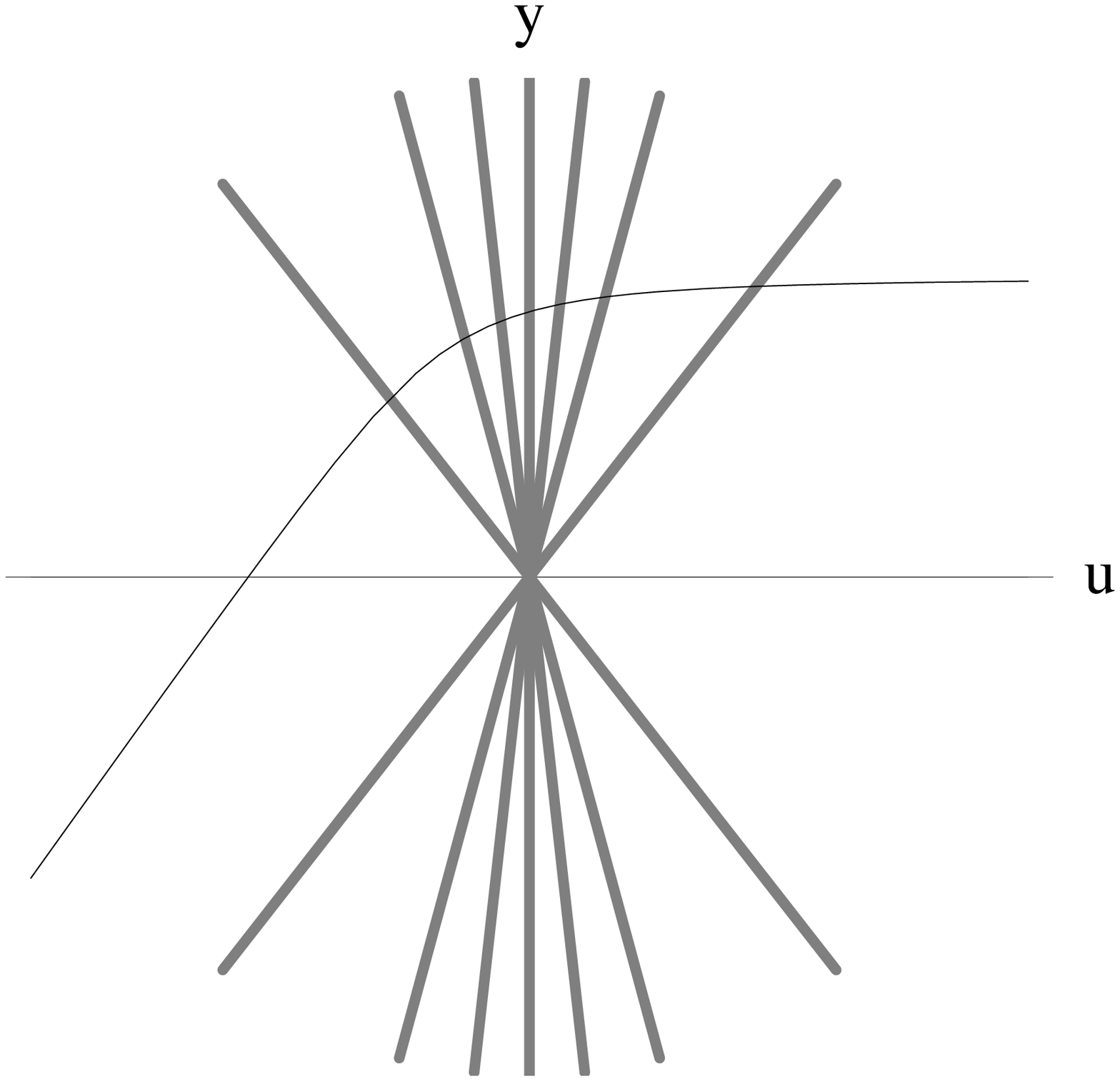}
\newline
\centering c)
\includegraphics[width=0.4\textwidth]{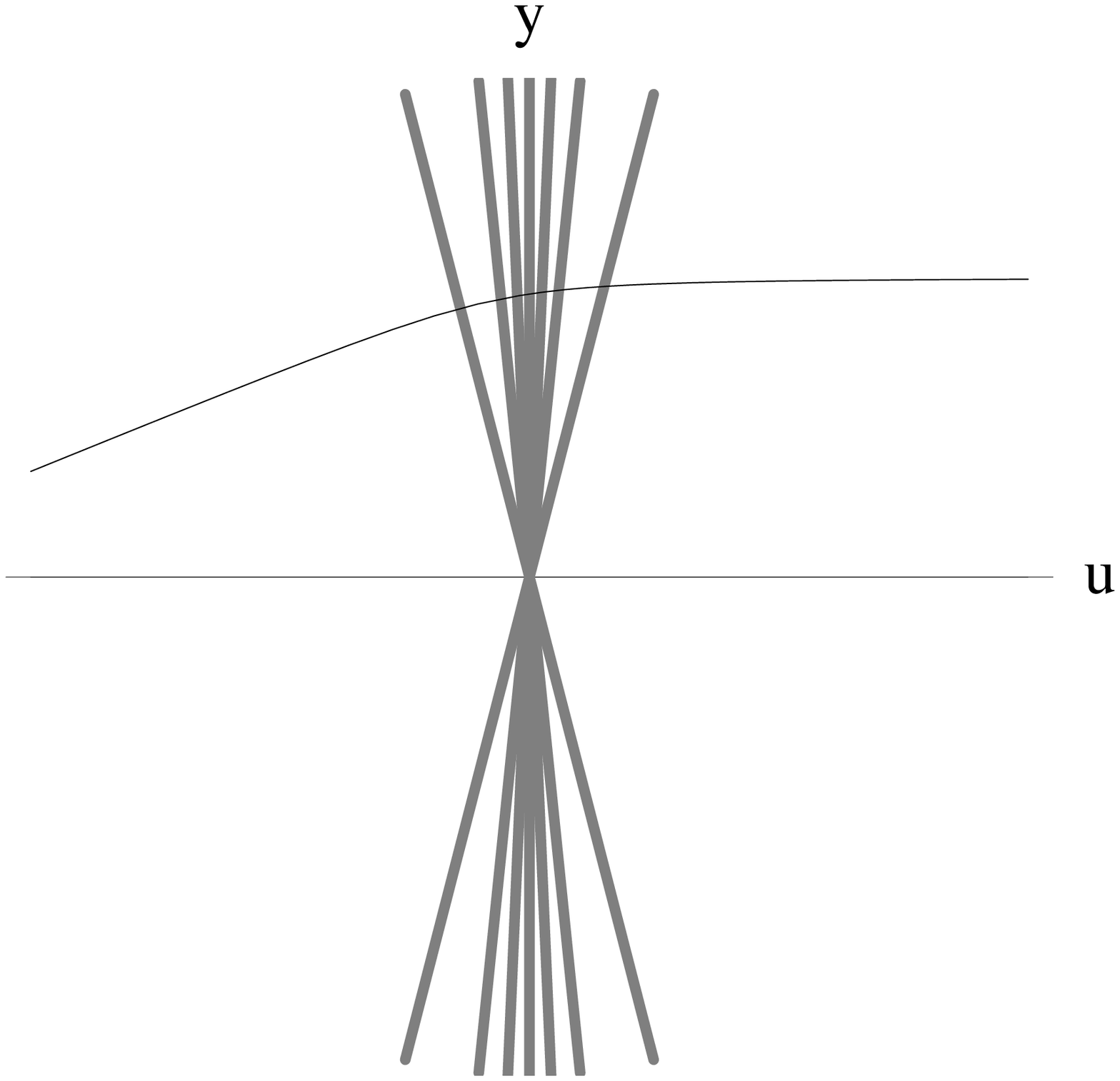}
\caption{Scattering trajectories of a slowly moving electric
charge by a) a stationary charged particle, b) a charged particle
moving at $v=0.94 \, c$, and c) a charged particle moving at
$v=0.99 \, c$. The gray lines indicate the electric field in each
case. With the moving particle going along the $z$ axis, the
coordinate $u=z- v t$, so this particle is always at $u=0$. The
charges are not held fixed here.} \label{fig:scattering}
\end{figure}

The situation we are considering is entirely analogous to the
electromagnetic scattering of a slowly moving charge by a rapidly
moving one as shown in Fig. \ref{fig:scattering}. When a charge is
highly boosted its electric field lines are compressed into the
directions transverse to its motion.  Most of the scattering of a
slow charge takes place while it moves through this region with a
more intense field. Away from it the slow charge experiences
almost no force.

Now consider what happens in gravitational scattering. In the
limit of an infinite boost all of the gravitational field is
confined inside a transverse plane, which is known as a
gravitational shock wave \cite{aichsexl}.  The metric around the
shock wave is just two patches of flat space sewn together, so it
is not surprising that the geodesics around a shock wave found by
Dray and 't Hooft \cite{thooft} correspond to particles travelling
freely except when they cross the shock wave front. The scattering
angle can easily be computed, but qualitatively we know that the
amount of bending will be proportional to the momentum of the
highly boosted particle since that is its gravitational `charge'.
Thus for a large enough boost we can get scattering at almost
right angle in the $u-y$ plane, and send a particle with a large
impact parameter onto a scattered trajectory that passes very
close to the path of the boosted particle.  If the scattered
particles end up within a distance smaller than the horizon of the
target particle then we certainly expect a black hole to form.

Let us now elaborate our argument. We will specialize to the case
of four dimensions since it illustrates the issues in the simplest
way possible. Consider a collision of two massive particles with
rest masses $m$ and $M$, which move towards each other with
relative (and relativistic!) velocity $\vec {\rm v}$, and impact
parameter ${ b}$. When they are far away from each other, as long
as they are massive, we can always go to the rest frame of one of
them, say $M$. At large relative velocities, ${\rm v} \rightarrow
1$, the gravitational field of the particle $m$ is extremely
strongly boosted in the rest frame of the particle $M$. Most of
the field is being carried by the particle momentum $p \sim E =
{m}/{\sqrt{1-{\rm v}^2}}$, by the equivalence principle. In this
limit, we can approximate the field by the linearized
Schwarzschild solution, boosted to a very large velocity $\vec
{\rm v}$. The nonlinear corrections to the field can be organized
as an expansion in $m/p$, such that in the infinite boost limit,
where we also take $m \rightarrow 0$ and hold $p$ fixed, the
metric reduces to an exact shock wave metric. In $4D$, the shock
wave fields are given in \cite{aichsexl,thooft}, and can be
generalized beyond $4D$ \cite{higherdwaves} and to codimension-2
tensional branes \cite{kalkil}.

In the infinite boost limit, when the source has zero rest mass,
the shock wave field is completely confined in the null transverse
surface orthogonal to the direction of motion, anchored at the
instantaneous location of the particle. For fast particles of
nonzero rest mass, the shock wave approximation breaks down far
away from the moving particle, because the field lines will spread
out of this surface. This will occur at transverse distances from
the source which are of the order of $\ell \sim
{r_h(m)}/{\sqrt{1-{\rm v}^2}}$.  Beyond this distance, we can't
neglect the spreading of the field lines in the direction of
motion, because an observer farther away than $\ell$ will see the
effects of retardation due to the slightly sub-luminal motion of
the source in the limit ${\rm v} \rightarrow 1$. Conversely,
whenever the impact parameter ${ b}$ of the collision is smaller
than $\ell$, we can use the shock wave field to extract the
information about the black hole formation to the leading order in
$m/p$.

We can estimate how far from the source we can trust the shock
wave approximation in practice. First, we have $1/\sqrt{1-{\rm
v}^2} \sim p/m$. Next, the gravitational radius $r_h$ of a
particle and its mass are related through the Schwarzschild
solution by $r_h = 2 G_N m$. Using this, the scale $\ell$ out to
which we can use the shock wave approximation for the field of the
fast particle is
\beq \ell \sim G_N p \sim \frac{p}{8\pi M_4^2} \, ,
\label{skaleell}
\ee
where we have defined the $4D$ Planck mass $M_4^2 = \frac{1}{8\pi
G_N}$. Thus for impact parameters of the order of $b \ll \ell$ the
shock wave is an excellent approximation. Clearly, we also need to
require that ${ b} > 1/M_4$ in order to trust the classical
description in the first instance.

To write the shock wave solutions, we can use the simple
cutting-and-pasting technique of \cite{thooft}. Instead of the
lightcone coordinates, we will use the standard Minkowski
coordinates to simplify the presentation. We take the null line
$u=\frac{t-z}{2}=0$ as the trajectory of the projectile, and
introduce a discontinuity in the orthogonal null coordinate $v = -
\frac{z+t}{2}$ by replacing $dv$ in the metric by $dv - f(x,y)
\delta(u) du$ \cite{thooft}. Here $f$ is the shock wave profile in
the transverse null plane anchored to the instantaneous location
of the particle, and $x$, $y$ denote the spatial dimensions along
it. The metric is, in Minkowski coordinates,
\begin{equation}
ds_4^2= dx^2 + dy^2 + dz^2 - dt^2 + 2 \delta(z-t) f(x,y) \bigl( dz
-dt \bigr)^2 \, . \label{shock}
\end{equation}
The shock wave profile $f(x,y)$ is the solution of the Poisson
equation in the transverse plane $\nabla_{2}^2 f = \frac{2p}{
M_4^{2} } \, \delta(x)\delta(y)$, which arises from imposing the
Israel junction condition on the wave front. This means that $f$
is the 2D Coulomb potential of a `charge' $p$ at the origin,
\be f = \frac{p}{\pi M_4^2}
\ln\Bigl(\frac{\sqrt{x^2+y^2}}{L}\Bigr) \, , \label{potential} \ee
where  $L$ is an arbitrary integration constant
\cite{aichsexl,thooft}.

The shock wave metric (\ref{shock}) shows that the gravitational
interaction -- as imparted by the field of the fast particle --
between our two colliding particles is practically instantaneous,
occurring at the moment when the shock front crosses the particle
$M$. Before and after this, the particle $M$ moves inertially.
Before the collision its gravitational field will deflect the
projectile, but this deflection angle, $\Delta \simeq \frac{G_N
M}{ b}$, will be sufficiently small to be neglected when ${ b} \gg
G_N M$. This justifies our choice of the initial rest frame of $M$
as the stage on which to set up our analysis, and our neglecting
the field of $M$ before the collision.

It is now straightforward to see what happens when a particle $M$
encounters the shock wave field (\ref{shock}). It will not remain
at rest, because at the moment the shock passes by it, it will  be
strongly attracted towards the source of the shock wave. As long
as the relative velocity $ {\rm v}$ is very close to the speed of
light, the interaction is impulsive, and the shock wave behaves as
a very thin {\it gravitational lens}. The total {\it physical}
deflection is Lorentz-invariant, and could be computed (with more
difficulty) in some other frame. The observed deflection angle,
however, needs to be properly transformed according to Lorentz
transformation rules, in order to account for relativistic
aberration between different inertial observers. It is now easy to
solve for the timelike geodesics of the shock metric (\ref{shock})
to find the spacetime trajectory of the particle $M$, at rest
before the collision. Since the problem is central, we can orient
it so that the motion is in the $x=0$ plane, by taking the initial
impact parameter to be along the $y$ direction. Then the
trajectory  of the target is
\be y = b -  \Theta(\tau) \,  {\rm v}_f \, \tau \, , ~~~~~ z =
\frac12 \, \Theta(\tau) \,  {\rm v}_f ^2 \,\tau - \Theta(\tau) \,
f(0,b) \, , ~~~~~ t = \tau +  \frac12 \, \Theta(\tau) \,  {\rm
v}_f ^2 \,\tau - \Theta(\tau) \, f(0,b) \, , \label{geodesics} \ee
where $\tau$ is the proper time of the particle $M$,
$\Theta(\tau)$ is the Heaviside step function,  $b$ is the impact
parameter, and ${\rm v}_f =
\partial_y f |_{y=b} = \frac{p}{\pi M_4^2 { b}} $ is the final
$4$-velocity component of $M$ in the $y$ direction. The collision
occurs at $\tau =0$, and the post-collision $4$-velocity is set by
the impulsive force pulling $M$ towards the projectile at the
moment of wave front crossing. Similarly, the projectile moves
along
\be y = 0 \, , ~~~~~~ z = t = \tau +  \frac12 \, \Theta(\tau) \,
{\rm v}_f ^2 \,\tau - \Theta(\tau) \, f(0,b) \, ,
 \label{geofast} \ee
where the parameterization of the solution is chosen to
synchronize the projectile with the target clock before the
collision.

Now, in the frame where it was initially at rest, the target
particle $M$ after the collision may end up moving {\it very
fast}. This will happen for impact parameters $b \ll p/M_4^2$. For
greater impact parameters, the target will move slowly, and
nothing dramatic will happen, but as we will see that case is
irrelevant for black hole formation.  We can find the
post-collision coordinate velocities of the mass $M$ in the $y$
and $z$ directions from the post-collision trajectory
(\ref{geodesics}),
\be {\rm v}_y = \frac{dy}{dt} = - \frac{{\rm v}_f}{1+\frac{{\rm
v}_f^2}{2}} \, , ~~~~~~~~~~~~  {\rm v}_z = \frac{dz}{dt} =
\frac{\frac{{\rm v}^2_f}{2} }{1+\frac{{\rm v}_f^2}{2}} \, ,
\label{velocities} \ee
so that after the collision, the speed of $M$ in the frame where
it was originally at rest is $ {\rm v}_M =  \sqrt{{\rm v}_y^2 +
{\rm v}_z^2} \rightarrow 1$, when ${\rm v_f} = \frac{p}{\pi M_4^2
b} \gg 1$. However, Eq. (\ref{velocities}) reveals that this is
because after crossing the shock wave, the target began moving
very rapidly in the $z$-direction, chasing after the projectile.
The velocity of the target particle $M$ in the $y$ direction
remains very small. The shock pulls the target along, accelerating
it to a very high velocity almost parallel with the shock source.

To see what happens next, we switch to a new inertial frame,
boosting in the $z$ direction to cancel the ${\rm v}_z$ component
of  the target post-collision velocity. We can always do this
after the collision, since the shock wave field has moved ahead
and the target is chasing after it in the vacuum left in the
projectile's wake. In this new frame the target moves very slowly
in the $y$ direction, and the spatial separation between the two
particles is $r^2 \simeq (\Delta y)^2 + \tau^2$, where  $\Delta y$
is the transverse distance between the particle trajectories after
the collision and $\tau$ is the distance that the fast particle
travelled after the collision, moving at near the speed of light.
This is because $\tau$ is the proper time of the target, as can be
checked from the solution (\ref{geodesics}), and since the target
is moving very slowly the difference between the coordinate time
and proper time is minuscule. Using our solutions and the
expression for $ {\rm v}_f$, this yields $r^2 \simeq { b}^2 (1-
\frac{p}{\pi M_4^2 { b}} \frac{\tau}{ b})^2 + \tau^2$. It is
simple to see that this is minimized when $r^2 \simeq \tau^2_*$,
where the time
\be
\tau_* = \frac{\pi M_4^2 {b}^2}{p}
\label{timecols}
\ee
is the instant when the particle $M$ will pass through $x=y=0$.

While the target particle $M$ won't feel the gravitational effects
of the shock wave field any more, the projectile will have to work
against the gravitational attraction of the target in order to
continue fleeing away. If it is to escape away to infinity, it
must move faster than the escape velocity set by the field of the
target particle at a distance $\sim \tau_*$ from it. Since in this
frame the target is again slow, we can estimate its gravitational
pull on the projectile by its Newtonian potential at the minimal
separation between the particles after the collision. This yields
${V}_{\rm escape}^2 \simeq  \frac{M}{M_4^2 \tau_*}$ for the escape
velocity at a distance $r = \tau_*$ from $M$. Substituting Eq.
(\ref{timecols}) and noting that $M p \simeq { s}$, we finally
obtain
\be {V}_{\rm escape}^2 \simeq  \frac{ s }{M_4^{4} { b}^2}  \, .
\label{escape1} \ee
Now, exactly as Michell \cite{michell} and Laplace \cite{laplace}
first noted, the projectile will be unable to escape if ${V}_{\rm
escape} \ge 1$. Hence whenever the initial impact parameter
satisfies
\be
{ b} \le \frac{\sqrt{ s}}{M_4^2} \, ,
\label{impact}
\ee
the target will capture the projectile. In other words after being
focused by the shock wave the separation of the two particles is
$\tau_*<r_h(M)$, so a black hole must form. As a check note that
when $p \gg M$, $\sqrt{ s} \ll p$, and so indeed ${ b} \ll \ell$,
which therefore means that our approximations of the projectile
gravitational field by a shock wave are fully justified. Further,
in this limit ${\rm v}_f \gg 1$, so our boosting to the new
post-collision frame is consistent. It is also worth noting that
from Eq. (\ref{velocities}) $v_y = \pi \sqrt{s}/p \ll 1$ where we
have used the maximal value of $b$ from Eq. (\ref{impact}). Thus
even after crossing the shock wave the target particle is still
travelling slowly in the $y$ direction unless the initial impact
parameter was much smaller that the typical value for black hole
formation. Moreover, imposing ${ b} > 1/M_4$ we find from Eq.
(\ref{impact}) that for black hole to form we must ensure ${ s} >
M_4$. Hence the center-of-mass energy for black hole formation
must be trans-Planckian. Then, the cross section for the black
hole to form is given by the final black hole area,
\be \sigma_{\rm black~hole} \simeq \pi { b}^2_{ critical} \simeq
\frac{ s}{M_4^4} \, . \ee
This completes our argument.

We conclude that indeed the estimates of the black hole production
cross section by the classical center-of-mass horizon area are
justified. In a frame where one of the particles starts at rest,
the process of black hole formation looks like the capture of the
projectile by the target particle, after the target is focused
down to a very small impact parameter by the shock wave field of
the projectile. For precision calculations, one may still need to
use the more involved derivations of the formation of trapped
surfaces in the collision of two shock waves. It would, however,
be interesting to refine our argument and see how the results of a
next-to-leading order treatment of the classical capture compare to
the trapped surface formation criteria.

It would also be interesting to extend this reasoning to
higher-dimensional examples and braneworld setups \cite{work},
which could be relevant for the forthcoming LHC searches.

\bigskip

We thank  M. Park for useful discussions. NK and JT were supported
in part by the DOE Grant DE-FG03-91ER40674. NK was also supported
in part by a Research Innovation Award from the Research
Corporation.



\begin{thebibliography}{99}

\bibitem{add}
N.~Arkani-Hamed, S.~Dimopoulos and G.~R.~Dvali,
Phys.\ Lett.\  B {\bf 429}, 263 (1998).

\bibitem{bhlhc}
P.~C.~Argyres, S.~Dimopoulos and J.~March-Russell,
Phys.\ Lett.\ B {\bf 441}, 96 (1998);
T.~Banks and W.~Fischler,
{\tt hep-th/9906038};
R.~Emparan, G.~T.~Horowitz and R.~C.~Myers,
Phys.\ Rev.\ Lett.\  {\bf 85}, 499 (2000);
S.~Dimopoulos and G.~Landsberg,
Phys.\ Rev.\ Lett.\  {\bf 87}, 161602 (2001);
S.~B.~Giddings and S.~D.~Thomas,
Phys.\ Rev.\ D {\bf 65}, 056010 (2002).

\bibitem{penrose}
R.~Penrose, unpublished, 1974.

\bibitem{deathpayne}
P.~D.~D'Eath and P.~N.~Payne,
Phys.\ Rev.\ D {\bf 46}, 658 (1992);
Phys.\ Rev.\  D {\bf 46}, 675 (1992);
Phys.\ Rev.\  D {\bf 46}, 694 (1992).

\bibitem{eardley}
D.~M.~Eardley and S.~B.~Giddings,
Phys.\ Rev.\ D {\bf 66}, 044011 (2002).

\bibitem{voloshin}
M.~B.~Voloshin,
Phys.\ Lett.\  B {\bf 518}, 137 (2001);
Phys.\ Lett.\  B {\bf 524}, 376 (2002)
[Erratum-ibid.\  B {\bf 605}, 426 (2005)].

\bibitem{michell}
J. Michell, Phil. Trans. Roy. Soc., {\bf 74}, 35-57 (1784).

\bibitem{laplace}
Pierre-Simon, Marquis de Laplace, {\it Exposition du syst\`eme du Monde}, 1796.

\bibitem{aichsexl}
P.~C.~Aichelburg and R.~U.~Sexl,
Gen.\ Rel.\ Grav.\ {\bf 2}, 303 (1971).

\bibitem{thooft}
T.~Dray and G.~'t Hooft,
Nucl.\ Phys.\ B {\bf 253}, 173 (1985);
Class.\ Quant.\ Grav.\  {\bf 3}, 825  (1986).

\bibitem{higherdwaves}
V.~Ferrari, P.~Pendenza and G.~Veneziano,
Gen.\ Rel.\ Grav.\  {\bf 20}, 1185 (1988);
H.~de Vega and N.~Sanchez,
Nucl.\ Ph.\ B {\bf 317}, 706 (1989).

\bibitem{kalkil}
N.~Kaloper and D.~Kiley,
JHEP {\bf 0603}, 077 (2006).


\bibitem{work}
N.~Kaloper and J.~Terning,
work in progress.

\end{thebibliography}
\end{document}